\documentclass{PoS}

\title{Oscillons and oscillating kinks in the Abelian-Higgs model}

\ShortTitle{Oscillons and oscillating kinks in the Abelian-Higgs model}

\author{\speaker{Charilaos Tsagkarakis}\\
        Department of Physics, University of Athens\\
        E-mail: \email{ctsagkarakis@phys.uoa.gr}}

\author{V. Achilleos\\
       Department of Physics, University of Athens\\  
        E-mail: \email{vachill@phys.uoa.gr.}}
\author{F.K. Diakonos\\
        Department of Physics, University of Athens\\  
        E-mail: \email{fdiakono@phys.uoa.gr}}
\author{D.J. Frantzeskakis\\
        Department of Physics, University of Athens\\  
        E-mail: \email{dfrantz@phys.uoa.gr}}
\author{G.C. Katsimiga\\
        Department of Physics, University of Athens\\  
        E-mail: \email{liakatsim@gmail.com}}
\author{X.N. Maintas\\
        Department of Physics, University of Athens\\  
        E-mail: \email{xmaintas@phys.uoa.gr}} 
\author{E. Manousakis\\
        Department of Physics, University of Athens\\
        Department of Physics, Florida State University, Tallahassee, 
Florida, USA\\  
        E-mail: \email{emanous@phys.uoa.gr}}    
\author{A. Tsapalis\\
        Department of Physics, University of Athens\\  
        Hellenic Naval Academy, Greece\\
        E-mail: \email{tsapalis@snd.edu.gr}}   

\abstract{
 We study the classical dynamics of the Abelian Higgs model 
employing an asymptotic multiscale expansion method, which  
uses the ratio of the Higgs to the gauge field amplitudes as a small parameter.  
We derive an effective 
nonlinear Schr\"{o}dinger equation for the gauge field, 
and a linear equation for the scalar field containing the gauge field as a nonlinear source.  
This equation is used to 
predict the existence of oscillons and oscillating kinks for certain regimes of the ratio of the 
Higgs to the gauge field masses. Results of  numerical simulations are found to be in very good agreement with 
the analytical findings, and show that the oscillons are robust, while kinks are unstable. 
It is also demonstrated that oscillons emerge spontaneously as a result of the onset of the modulational instability 
of plane wave solutions of the model. Connections of the obtained solutions with 
the phenomenology of superconductors is discussed.}

\FullConference{Proceedings of the Corfu Summer Institute 2014 \\
                 3-21 September 2014\\
                 Corfu, Greece}

\begin{document}

\section{Introduction}

Soliton solutions of field theoritical models play a significant role in the description of physical phenonena occuring in a wide class of systems ranging from high energy physics to cold atoms and granular media \cite{earlyuniverse,amin,stama068,rajaraman,granular,water,sinegordon,bogo,campbell,gleiser94,honda,salmi06,salmi08,fodor06,fodor08,
gleiser08,segur,gleiser2,farhi05,graham,sfakia,bec,spiel1,bholes,emergent}. In particular classical solutions of the mean field  Gor'kov - Eliashberg - Landau - Ginzburg theory are important for the phenomenology of superconductivity 
\cite{Gorkov,GorkElia,GinzbLand,Abrikosov} where the emerging effective field theory is the Abelian-Higgs model. Several solutions of the latter are found \cite{nielsen,gl,mclerran}.


In the present work \cite{presentwork}, we 
analyze the classical dynamics of the Abelian-Higgs model in $(1+1)$-dimensions by means of a multiscale 
expansion method \cite{multiplescales}.
We restrict our analysis on the case of small fluctuations of the condensate about its vacuum expectation value (vev). This is realized by choosing the amplitude of the Higgs field to be an order of magnitude smaller than that of the gauge field.  In this limit the dynamics simplify considerably and it is found that the scalar field performs asymmetric
oscillations around the classical vacuum. Such a scenario
occurs naturally considering the model just after the symmetry breaking i.e. close to the critical point. Then the minimum
of the scalar field potential is very flat and the asymmetric
cubic term is strong leading to an asymmetric shape of the
potential around it. However, results  are also found in the case where the gauge and the scalar field amplitudes are of the same order \cite{pre}. This scenario corresponds to a strong breaking of the underlying gauge
symmetry, which is far beyond the related critical point. In this case
the minimum of the potential occurs at the bottom of a deep
well, while the potential shape is almost symmetric around it
since the quadratic term dominates.

Employing the method of multiple scales (see also Ref.~\cite{emeis}), the original nonlinear coupled field equations are  
reduced to an effective nonlinear Schr\"{o}dinger (NLS) equation for the gauge field, and a linear equation for the Higgs field containing the gauge field as a source. These equations are analytically solved giving two types of localized solutions 
for the gauge field, namely in the form of oscillons and oscillating kinks.
Subsequently, we numerically integrate the original equations of motion using, as initial conditions, 
the analytically found solutions. 
We find that the analytical predictions are in excellent agreement with the numerical findings  for a large range of values of the parameters involved.  
We then discuss connections between the Abelian-Higgs model solutions 
and the phenomenology of superconducting materials, describing the temporal fluctuations of the condensate in a 
one-dimensional (1D) Josephson junction~\cite{paterno} due to the presence of oscillating (in time) magnetic and electric fields.

\section{The model and its analytical consideration}

\subsection{Formulation and Equations of Motion}

The $U(1)$-Higgs field dynamics is described by the Lagrangian:
\begin{eqnarray}
{\mathcal{L}}=-\frac{1}{4} F_{\mu \nu} F^{ \mu \nu} &+& (D_\mu \phi)^{*}(D^\mu \phi)
- V(\Phi^{*} \Phi),
\label{eq:eq1}
\end{eqnarray}
where $F_{\mu \nu}$ is the $U(1)$ field strength tensor, $D_\mu=\partial_{\mu} +ie \tilde{A}_{\mu}$ is the covariant 
derivative, and $V(\Phi^{*} \Phi)=\mu^2 \Phi^{*} \Phi + \lambda (\Phi^{*} \Phi)^2$ ($\lambda > 0$) 
is the Higgs self-interaction potential (asterisk denotes complex conjugate). In the broken phase, $\mu^2 <0$, 
a vev $\tilde{\upsilon }/ \sqrt{2}$ of the Higgs field arises classically: $\tilde{\upsilon}^2=-\mu^2 / \lambda$. 
We will focus on the dynamics of this system assuming that the Higgs field fluctuates slightly around its vev. 
In this case, we expand the field $\Phi$ as: $\Phi=\frac{1}{\sqrt{2}}(\tilde{\upsilon} +\tilde{H})$ 
and obtain the following equations of motion for $\tilde{A}_{\mu}$ and $\tilde{H}$:
%
\begin{eqnarray}
(\widetilde{\Box} &+& m_A^2) \tilde{A}_{\mu} - \partial_{\mu} (\partial_{\nu} \tilde{A}^{\nu}) + 2e^2 \tilde{\upsilon}  \tilde{H} \tilde{A}_{\mu} + {e^2}  \tilde{H}^2 \tilde{A}_{\mu} = 0,
\nonumber \\ \label{eq:eq2}\\
(\widetilde{\Box} &+& m_H^2 )\tilde{H} + 3 \lambda \tilde{\upsilon} \tilde{H}^2 + \lambda \tilde{H}^3 - {e^2} \tilde{A}_{\mu} \tilde{A}^{\mu} (\tilde{\upsilon} + \tilde{H}) = 0,
\nonumber \\
\label{eq:eq3}
\end{eqnarray}
%
where $m_A^2={e^2 \tilde{\upsilon}^2}$ and $m_H^2=2 \lambda \tilde{\upsilon}^2$.
We simplify Eqs.~(\ref{eq:eq2})-(\ref{eq:eq3}) choosing the field representation $\tilde{A}_0=\tilde{A}_1=\tilde{A}_3=0$, 
and $\tilde{A}_2=\tilde{A}\neq0$. Due to the fact that we are interested in 1D settings, 
the non vanishing $\tilde{A}_2$ field is a function solely of $x$ and $t$.
As a consequence, the Lorentz condition $\partial_{\nu} \tilde{A}^{\nu} = 0$ is trivially fulfilled and we are left with  a coupled system of equations for the gauge field 
$\tilde{A}(x,t)$ and the Higgs field. 
Furthermore, we write Eqs.~(\ref{eq:eq2})-(\ref{eq:eq3}) in a dimensionless form 
by rescaling the fields as: $\tilde{A}\rightarrow (m_A/e) A$, $\tilde{H} \rightarrow (m_A/e)H$, 
and space-time coordinates as: $\tilde{x}\rightarrow x/m_A$  and $\tilde{t}\rightarrow t/m_A$. 
Thus we end up with the following equations of motion:
\begin{eqnarray}
(\Box &+& 1)A + 2 A H + H^2 A = 0, \label{eq:eq4}\\
(\Box &+& q^2)H+\frac{1}{2}q^2 H^3 +\frac{3}{2}q^2 H^2 + H A^2 + A^2 = 0, 
\label{eq:eq5}
\end{eqnarray}
%
where parameter $q \equiv m_H/m_A$ is assumed to be of order $\mathcal{O}(1)$. 
The Hamiltonian corresponding to the above equations of motion is:
\begin{eqnarray}
\mathcal{H}= \frac{1}{2}\big[ (\partial_t A)^2 +  (\partial_x A)^2 + 
(\partial_t H)^2 + (\partial_x H)^2 \big] + V, 
\label{ham}
\end{eqnarray}
where indices denote partial derivatives 
with respect to  $x$ and $t$, and the potential $V$ is given by:
\begin{eqnarray}
V = \frac{q^2}{8}H^4+\frac{q^2}{2} H^3 + \frac{q^2}{2} H^2 + \frac{1}{2} H^2  A^2 + H A^2 + \frac{1}{2} A^2. \nonumber \\
\label{pot}
\end{eqnarray}
%
Notice that $V$ exhibits a single minimum at $(A,H)=(0,0)$.

\subsection{
Multiscale expansion and the NLS equation}

Considering a small fluctuating field $H$, 
we may employ a perturbation scheme, which uses a formal  
small parameter $0<\epsilon\ll 1$ 
defined by the ratio of the amplitude of the Higgs field to the amplitude of the gauge field, i.e., 
$\epsilon \sim H/A$. In particular, we will employ a multiscale expansion method 
\cite{multiplescales}, assuming that the Higgs and gauge fields depend on the set of independent variables 
$x_0=x$, $x_1=\epsilon x$, $x_2=\epsilon^2 x, \ldots$ and 
$t_0=t$, $t_1=\epsilon t$, $t_2=\epsilon^2 t, \ldots$.
Accordingly, the partial derivative operators 
are given (via the chain rule) by $\partial_x = \partial_{x_0} + \epsilon \partial_{x_1} + \ldots$, 
$\partial_t = \partial_{t_0} + \epsilon \partial_{t_1} + \ldots$.
Furthermore, taking into regard that the fields should be expanded around the trivial solution of 
Eqs.~(\ref{eq:eq4})-(\ref{eq:eq5}), as well as $H \sim \epsilon A$ as per our assumption, 
we introduce the following asymptotic expansions for $A$ and $H$: 
%
%
%
\begin{eqnarray}
A&=& \epsilon A^{(1)} + \epsilon^2 A^{(2)} + \ldots, \label{A}
\\
H&=& \epsilon^2 H^{(2)} + \epsilon^3 H^{(3)}+ \ldots,  \label{eq:eq6}
\end{eqnarray}
where $A^{(j)}$ and $H^{(j)}$ ($j=0,1,\ldots$) 
denote the fields at the order $\mathcal{O}(\epsilon^j)$. 
Substituting Eqs.~(\ref{A})-(\ref{eq:eq6}) into
Eqs.~(\ref{eq:eq4})-(\ref{eq:eq5}), and using the variables $x_0$, $x_1,\ldots$, $t_0$, $t_1,\ldots$, 
we obtain the following system of equations up to the third-order in the small parameter epsilon $\mathcal{O}(\epsilon^3)$:
\begin{eqnarray}
\!\!\!\!\!\!\!\!\!\!\!&\mathcal{O}(\epsilon)\!\!&:(\Box_0 + 1)A^{(1)}=0 ,
 \label{eq:order1} 
 \\
\!\!\!\!\!\!\!\!\!\!\!&\mathcal{O}(\epsilon^2)\!\!& : \left(\Box_0 +1 \right)A^{(2)} + 2(\partial_{t_0} \partial_{t_1}-\partial_{x_0} \partial_{x_1})A^{(1)}=0, \label{eq:order2A} 
\\
\!\!\!\!\!\!\!\!\!\!\!&\!\!&\;\;( \Box_0 + q^2)H^{(2)} + A^{(1)2}=0,  
\label{eq:order2H}
\\
\!\!\!\!\!\!\!\!\!\!\!&\mathcal{O}(\epsilon^3)\!\!&:
\left(\Box_0 + 1\right)A^{(3)} + 2\left(\partial_{t_0}\partial_{t_1}-\partial_{x_0} \partial_{x_1}\right)A^{(2)}
\nonumber
\\
\!\!\!\!\!\!\!\!\!\!\!&\!\!& +\left(\Box_1 + 2\partial_{t_0} \partial_{t_2}-2\partial_{x_0} \partial_{x_2} + 2H^{(2)}\right)A^{(1)}=0. 
 \label{eq:order3A}
\end{eqnarray}
The solution of Eq.~(\ref{eq:order1}) at the first order is: 
%
\begin{equation}
A^{(1)}=u(x_1,x_2,\ldots,t_2,t_3\ldots)e^{i\theta} + {\rm c.c.}, 
\label{eq:eq10}
\end{equation}
where $u$ is an unknown function, 
``c.c.'' stands for complex conjugate, while $\theta=Kx-\Omega t$, and the frequency $\Omega$
and wavenumber $K$ are connected 
through the dispersion relation $\Omega^2=K^2+1$. 
The solvability condition for the second-order equation~(\ref{eq:order2A}), is 
$(\partial_{t_0} \partial_{t_1}-\partial_{x_0} \partial_{x_1})A^{(1)}=0$ \footnote{The second term is ``secular'', 
i.e., is in resonance with the first term which becomes $\propto \theta \exp(i\theta)$, thus 
leading to blow-up of the solution for $t \rightarrow +\infty$.}. 
which satisfied in the frame moving with the group velocity  
 $\upsilon_g \equiv \partial{\Omega}/\partial{K}$, so that $u$ depends only on the 
variable $\tilde{x}_1=x_1-\upsilon_g t_1$ and $t_2$.
Then, the field $A^{(2)}$ is taken to be of the same functional form as $A^{(1)}$ [cf. Eq.~(\ref{eq:eq10})].
Taking into account the above results, Eq.~(\ref{eq:order2H}) for the Higgs field becomes: 
%
\begin{equation}
H^{(2)}=B\left(b \vert u \vert^2 + u^2 e^{-2 it} +{\rm c.c.}
\right),
\label{eq:eq11}
\end{equation}
where $B = 1/(4-q^2)$ and $b=-2/Bq^2$; here, it is assumed that $q \ne 0$ and $q \ne 2$, as for 
these limiting values the field $H^{(2)}$ (which is assumed to be of $\mathcal{O}(1)$ for the perturbation expansion to be valid) 
becomes infinitely large. 
Finally, at the order $\mathcal{O}(\epsilon^3)$, we first note that the second term 
vanishes  in the frame moving with $\upsilon$. Thus, the solvability condition 
(obtained ---as before--- by requiring that the secular part is zero) becomes:  
$\left(\Box_1 + 2\partial_{t_0} \partial_{t_2} -2\partial_{x_0} \partial_{x_2}+ 2H^{(2)}\right)A^{(1)}=0$. 
In the following, for simplicity, we will consider the zero momentum case 
($K=0$); 
hence, $\upsilon_g=0$, and also $\tilde{x}_1=x_1$, i.e., the field $A_1$ is at rest. 
In this case, we derive the following NLS equation
for the unknown function $u(x_1,t_2)$:
\begin{eqnarray}
i \partial_{t_2} u +\frac{1}{2} \partial_{x_1}^2 u  + s \vert u\vert^2 u = 0,\label{eq:eq12}
\end{eqnarray}
where the parameter $s$ is given by
\begin{eqnarray}
s &=& \frac{2}{q^2} + \frac{1}{q^2-4}. \label{s}
\end{eqnarray}

The above NLS equation possesses exact localized solutions of two different types, 
depending on the sign of $s$: for $s>0$, the solutions are sech-shaped solitons, while for 
$s<0$ the solutions are tanh-shaped kinks. These solutions have been extensively
studied in a variety of physical contexts including nonlinear optics \cite{yuri}, 
water waves \cite{johnson}, atomic Bose-Einstein condensates \cite{emergent}, and so on. 
Here, the type of solution, i.e., the sign of the nonlinear term in Eq.~(\ref{eq:eq12}), 
depends solely on the parameter $q$.
%

In particular, for $q<1.63$ and $q>2$, we obtain $s>0$ [attractive (focusing) nonlinearity] and the 
soliton solution of Eq.~(\ref{eq:eq12}) has the form:
\begin{equation}
u = u_0 {\rm sech} \left(\sqrt{s} u_{0} x_1 \right) e^{-i\omega t_2}, \label{eq:eq13}
\end{equation}
where $u_0$ is a free parameter [considered to be of order $\mathcal{O}(1)$] characterizing 
the amplitude of the soliton, while the soliton frequency is $\omega = -(1/2) u^2_{0}s$.
Accordingly, the approximate solutions of Eqs.~(\ref{eq:eq4})-(\ref{eq:eq5}) for $A$ and $H$ can be expressed 
in terms of the original variables $x$ and $t$ as follows:
\begin{eqnarray}
\!\!\!\!
A &\approx & 2\epsilon u_{0} {\rm sech}\left(\epsilon u_{0} \sqrt{s} x \right) 
\cos\left[\left(1-\frac{1}{2}(\epsilon u_0)^2 s\right)t \right], 
\label{eq:eq14} \\
\!\!\!\!
H &\approx & (\epsilon u_0)^2 B{\rm sech}^2\left(\epsilon u_{0} \sqrt{s}  x \right)  
\left\{ b+2\cos \left[2\left(1-\frac{1}{2}(\epsilon u_0)^2 s \right) t \right]\right\},  
\label{eq:eq15}
\end{eqnarray}
characterized by the single free parameter  $\epsilon u_{0}$.
The above approximate solutions for the gauge field $A$ and the
Higgs field $H$ are localized in space (decaying for $|x|\rightarrow \infty$) and are oscillating in time; thus, they correspond to 
{\it oscillons} (alias breathers). At the leading-order in $\epsilon$, the oscillation frequencies for $A$ and $H$ are given by: 
$\omega_A\approx 1$ and $\omega_H\approx 2$ (in units of $m_A$) respectively.

On the other hand, for $1.63<q<2$, we obtain $s<0$ [repulsive (defocusing) nonlinearity], 
so that Eq.~(\ref{eq:eq12}) possesses 
a kink solution of the following form:
\begin{equation}
u= u_0 \tanh \left( \sqrt{|s|}u_{0} x_1 \right) e^{-i\omega t_2}, \label{eq:eq16}
\end{equation}
where $u_0$ is a $\mathcal{O}(1)$ free parameter and $\omega=u^2_0 |s|$. In this case, the approximate solutions 
for the gauge and Higgs fields read:
\begin{eqnarray}
\!\!\!\!\!\!
A &\approx & 2 \epsilon u_{0} \tanh \left(\sqrt{|s|}\epsilon u_0 x \right) 
\cos\left[\left(1+(\epsilon u_{0})^2 |s|\right)t\right], \label{eq:eq17} \\
\!\!\!\!\!\!
H &\approx & (\epsilon u_{0})^2 B \tanh^2 \left(\sqrt{|s|} \epsilon u_{0} x \right)  
\left\{b+2\cos \left[2\left(1+ \epsilon^2 u^2_{0}|s|\right)t\right]\right\}.
 \label{eq:eq18}
\end{eqnarray}
These solutions correspond to oscillating kinks, for both fields, which are zero exactly at the core of the kink
and acquire non vanishing values at $|x|\rightarrow \infty$. Again the fields oscillate with the frequencies 
$\omega_A\approx 1$ and $\omega_H\approx 2$.


\section{Numerical results}
To further elaborate on our analytical findings, in this Section we will present results stemming from  
numerical integration of Eqs.~(\ref{eq:eq4})-(\ref{eq:eq5}). Our aim is to check the validity of our analytical 
predictions, namely the existence of oscillons and oscillating kinks, as well as study numerically their stability. 
The equations of motion are integrated by using a fourth-order Runge-Kutta time integrator and 
a pseudo-spectral method for the space derivatives \cite{jianke}. The lattice spacing used 
was fixed to $\Delta x= 0.2$, the time step $\Delta t=10^{-2}$, and the total length of the lattice $L=400$.
In all simulations we used, as an initial condition, the approximate solutions of Eqs.~(\ref{eq:eq14})-(\ref{eq:eq15}) for the oscillons and of Eqs.~(\ref{eq:eq17})-(\ref{eq:eq18}) for the oscillating kinks for $t=0$; 
in all cases we have fixed the free parameter value to $u_0=1$, and the ratio of the two fields amplitudes 
to $\epsilon=0.1$. 

\subsection{Oscillons}
\begin{figure}[tbp]
\center
\includegraphics[scale=0.4]{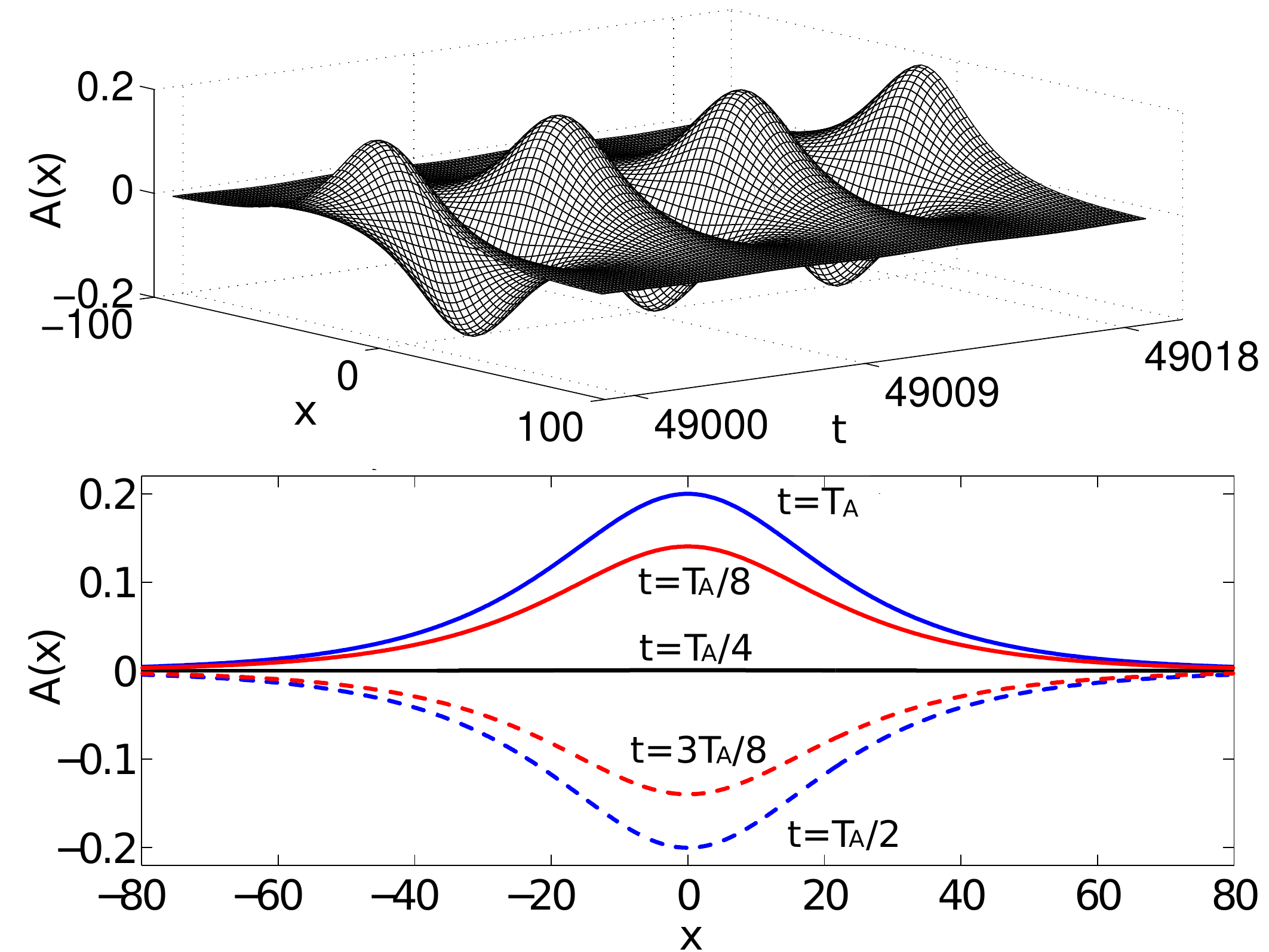}
\caption{(Color online) Top panel: Evolution of the gauge field $A$ as a function of $x$ and $t$. Bottom panel: Profile snapshots of the field at different instants of its breathing motion. Parameter values used are $\epsilon=0.1$, $q=1.5$ and the oscillation period is $T_A=2\pi$. }
\label{fig1}
\end{figure}
\begin{figure}[tbp]
\center
\includegraphics[scale=0.4]{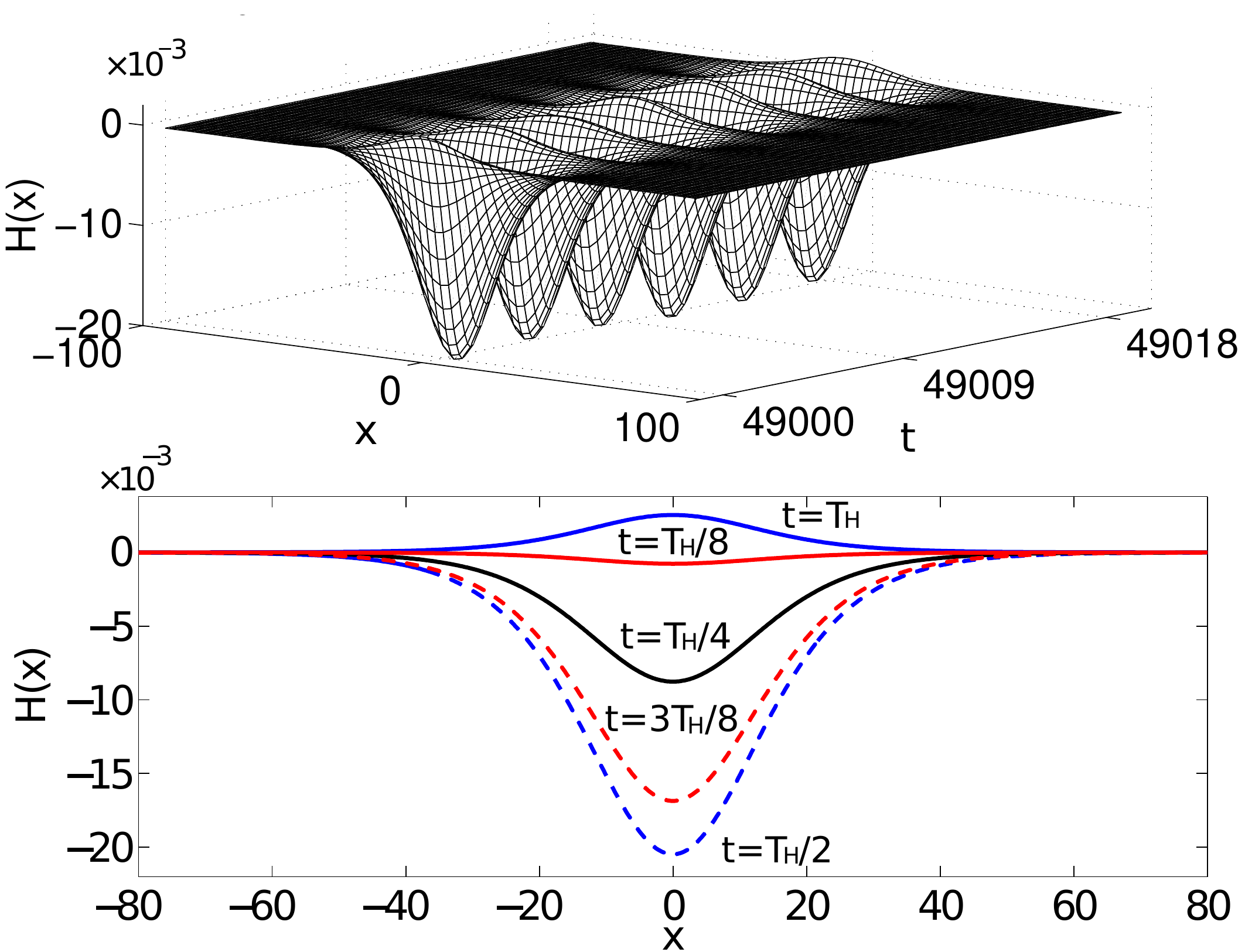}
\caption{(Color online) Same as in Fig.~1, but for the 
Higgs field $H$. Notice that the oscilation period of the Higgs field is $T_H=\pi$.}

\label{fig2}
\end{figure}
\begin{figure}[tbp]
\center
\includegraphics[scale=0.4]{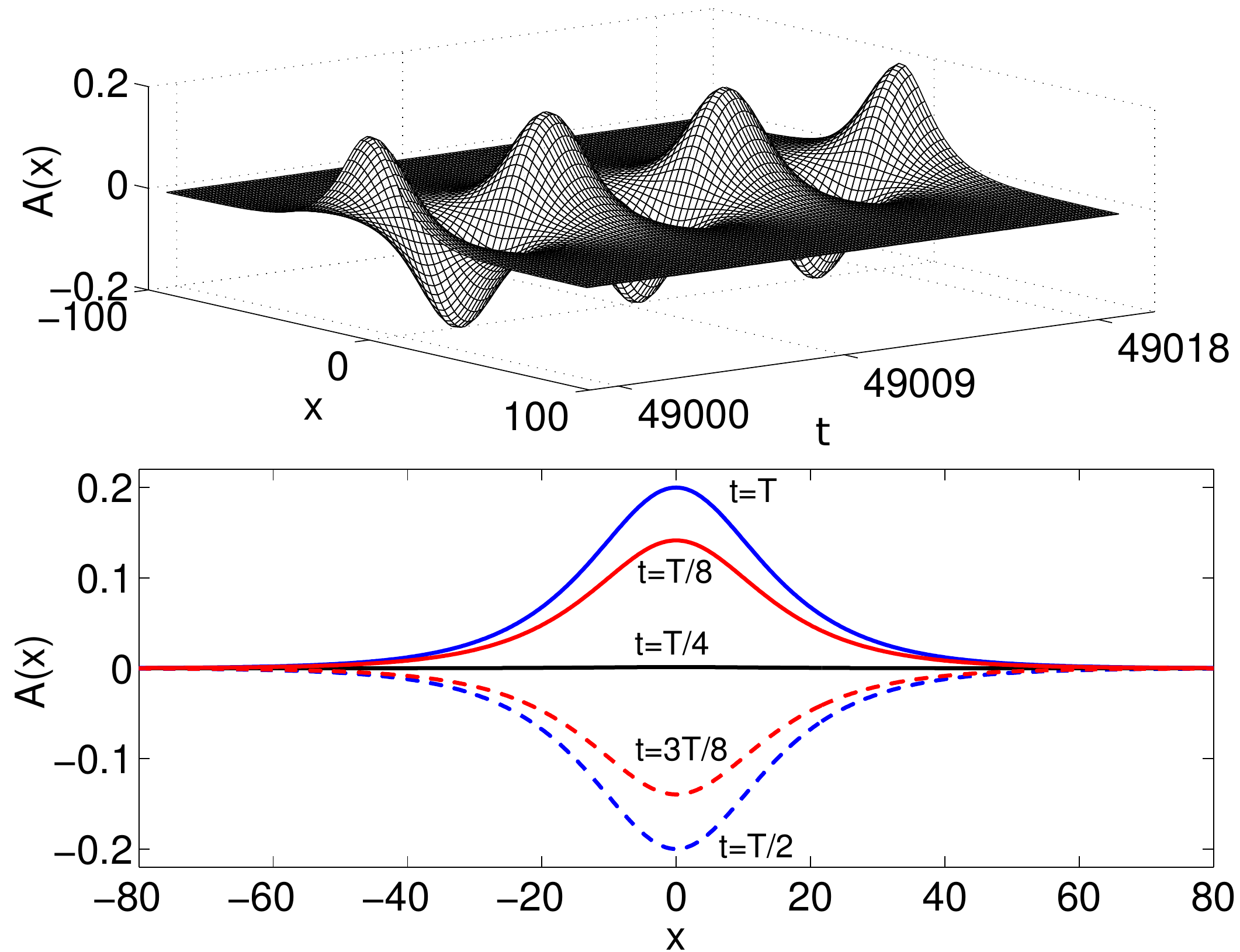}
\caption{(Color online) Same as Fig.~1, for parameter values $\epsilon=0.1$ and $q=2.5$. }
\label{fig3}
\end{figure}
\begin{figure}[t]
\center
\includegraphics[scale=0.4]{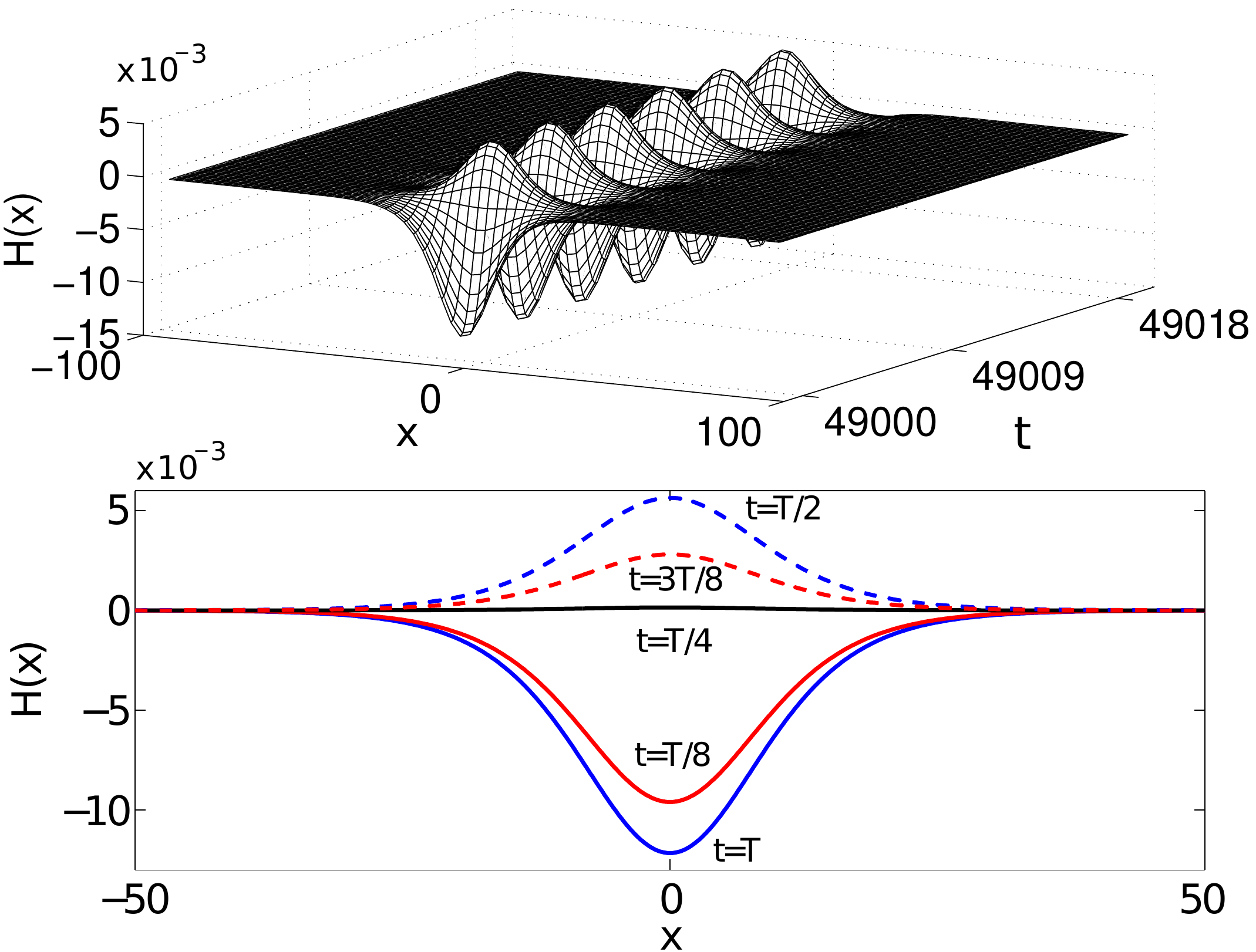}
\caption{(Color online) Same as Fig.~2, with parameter values $\epsilon=0.1$, $q=2.5$.}
\label{fig4}
\end{figure}

In the top panels of Fig.~\ref{fig1} and Fig.~\ref{fig2} a contour plot showing the evolution of an 
oscillon solution is shown (for $q=1.5$) at the end of the integration time 
$t=5\times 10^{4}$. Note that the periods of
the two fields are $T_A\equiv 2\pi/\omega_A=2\pi$ and $T_H\equiv 2\pi/\omega_H=\pi$, and thus $3$ oscillations for the
gauge field and $6$ oscillations for the $H$ field are depicted.
In the bottom panels of Figs.~\ref{fig1} and \ref{fig2}, we show the profiles of $A$ and $H$ ,
during the interval of one period of exhibiting their breathing motion. While the gauge field $A$ performs symmetric
oscillations, the Higgs field oscillates asymmetrically with respect to its equilibrium point. This is due to the constant $b$
in the solution of Eq.~(\ref{eq:eq15}).  
The evolution of an oscillon for 
$q=2.5$ (lying in the second region where oscillons exist) 
is also shown in the top panels of Fig.~\ref{fig3} and Fig.~\ref{fig4}. The
breathers become more localized in this region, as can be seen from their snapshots (bottom panels of
Figs.~\ref{fig3}-\ref{fig4}).

\begin{figure}[t]
\center
\includegraphics[scale=0.45]{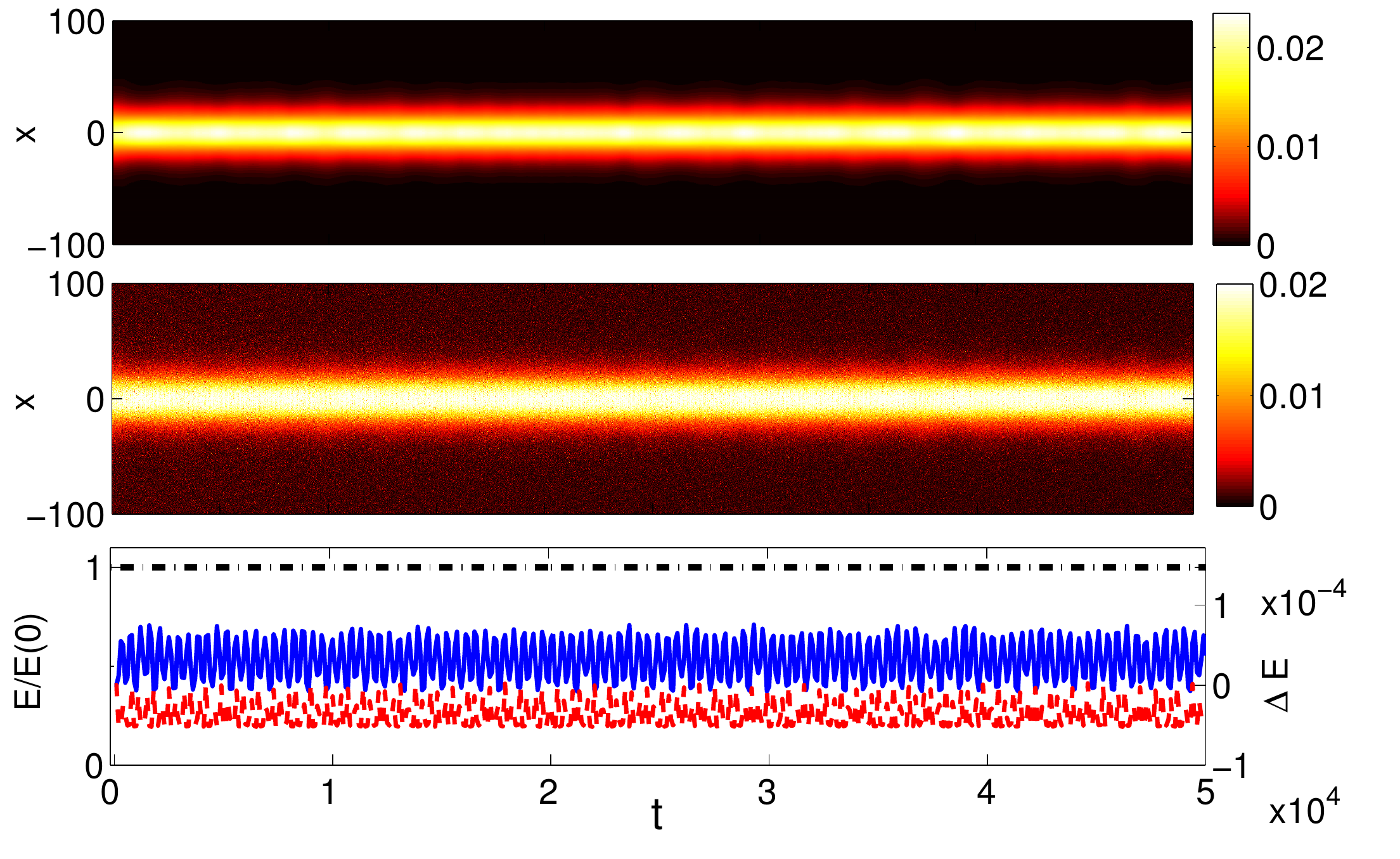}
\caption{(Color online) Top and middle panel: contour plot showing the energy density $\mathcal{H}(t)$, as 
a function of time, for $q=1.5$. 
For the middle panel the initial condition was perturbed by a small random noise with an amplitude $10\%$ of 
the oscillon's amplitude. Bottom panel: the total energy normalized to its initial value at $t=0$ 
dashed dotted line [upper (black)] .
The energy difference $\Delta E$, for $q=1.5$  solid [middle (blue)] line and $q=2.5$ dashed [bottom (red)] line.}
\label{fig5}
\end{figure}

In order to demonstrate the localization of the energy of the oscillons, in the top and middle panels of Fig.~\ref{fig5},
we show a contour of the total Hamiltonian density [cf. Eq.~(\ref{ham})] for $q=1.5$. 
The top panel, corresponds to an initial condition using only Eqs.~(\ref{eq:eq14})-(\ref{eq:eq15}), while in the middle
panel we have also added a random Gaussian noise to the initial condition, as large as $10\%$ of the
oscillon's amplitude (the rest of the parameters are as in Fig.~\ref{fig1}).
We observe that the energy density remains localized during the numerical integration in both cases
(with and without noise).
We would like to stress here that, the latter result, shows that oscillons are robust
under the effect of a random noise, and this was also confirmed for different values of $q$ in 
the domain of existence of the oscillon. 
For completeness, the total energy $E(t)=\int^{\infty}_{-\infty} \mathcal{H}dx$, normalized to
its initial value $E(0)$, is also shown (dashed dotted line)in the bottom panel of Fig.~\ref{fig5}. In fact the energy fluctuations, 
defined as $\Delta E = E - E(0)$, 
as shown in Fig.~\ref{fig5} [for $q=1.5$ (top solid line) and $q=2.5$ (bottom dashed line)]
are of the order of $10^{-4}$.


\subsection{Spontaneous oscillon formation}

In this section we show that oscillons can {\it emerge spontaneously}, through the mechanism of 
modulation instability (for more details see Ref.~\cite{mi}). The 
latter concerns the instability of 
plane wave solutions of the NLS Eq.~(\ref{eq:eq12}), 
of the form
\begin{equation}
u(x_1, t_2)= u_0 \exp{[i\left(kx_1 - \omega t_2\right)]}, \label{plane}
\end{equation}
under small perturbations. 
To briefly describe the emergence of this instability, we consider
the following ansatz:
\begin{eqnarray}
u=(u_0+ W) \exp{[i\left(kx_1 - \omega t_2\right)+i\Theta)]}, 
\label{miansatz1} 
\end{eqnarray}
where the amplitude and phase perturbations $W$ and $\Theta$ are given by:
\begin{eqnarray}
W&=& \varepsilon W_0 \exp{[i\left(Q x_1 - P t_2\right)]} + {\rm c.c.},
 \label{du}
 \\ 
\Theta &=&\varepsilon \Theta_0 \exp{[i\left(Q x_1 - P t_2\right)]}+ {\rm c.c.}, 
\label{dth}
\end{eqnarray}
with $W_0$ and $\Theta_0$ being constants, 
and $\varepsilon$ 
being a formal small parameter.
Substituting Eqs.~(\ref{miansatz1})-(\ref{dth}) into Eq.~(\ref{eq:eq12}) 
it is found that, for $k=0$, the frequency $P$ and the wavenumber $Q$ 
of the perturbations, obey the dispersion relation:
\begin{eqnarray}
P^2=(1/2)|Q|^2(|Q|^2/2-2s|u_0|^2). 
\label{dispmi}
\end{eqnarray}
It is evident from Eq.~(\ref{dispmi}) that for $s>0$ there exists an instability band 
for wavenumbers $ Q^2< 4u_0^2 s$, where $P$ becomes complex. Note that this instability can only 
emerge in the region where oscillons exist. 
When this instability manifests itself, the exponential growth of the
perturbations leads to the generation of localized excitations, which are identified as the oscillons 
described in Eqs.~(\ref{eq:eq14})-(\ref{eq:eq15}). 

\begin{figure}[t]
\center
\includegraphics[scale=0.48]{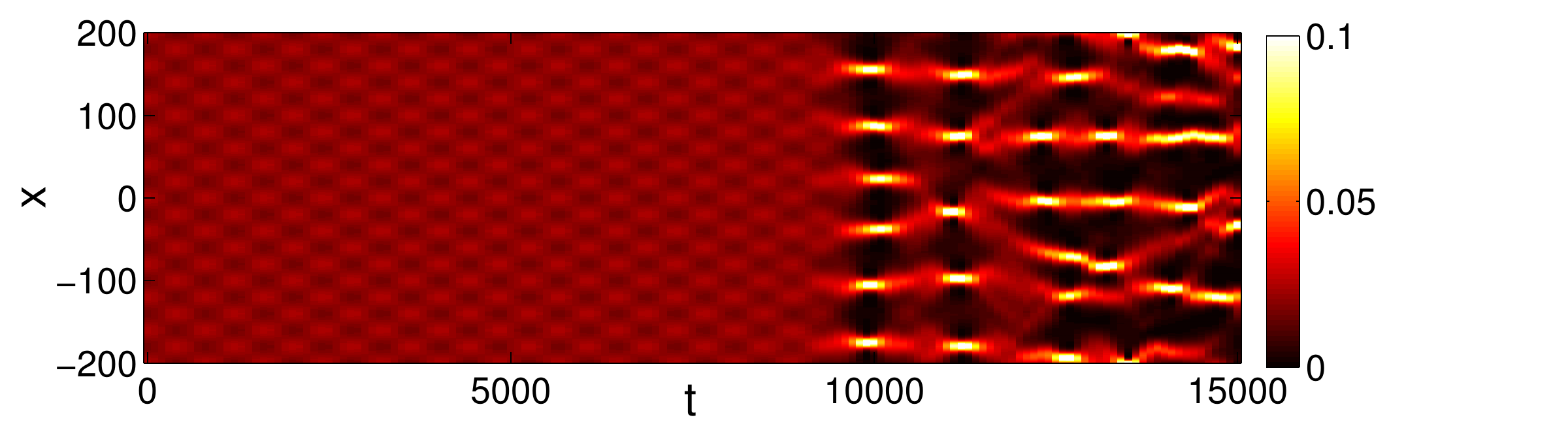}
\includegraphics[scale=0.5]{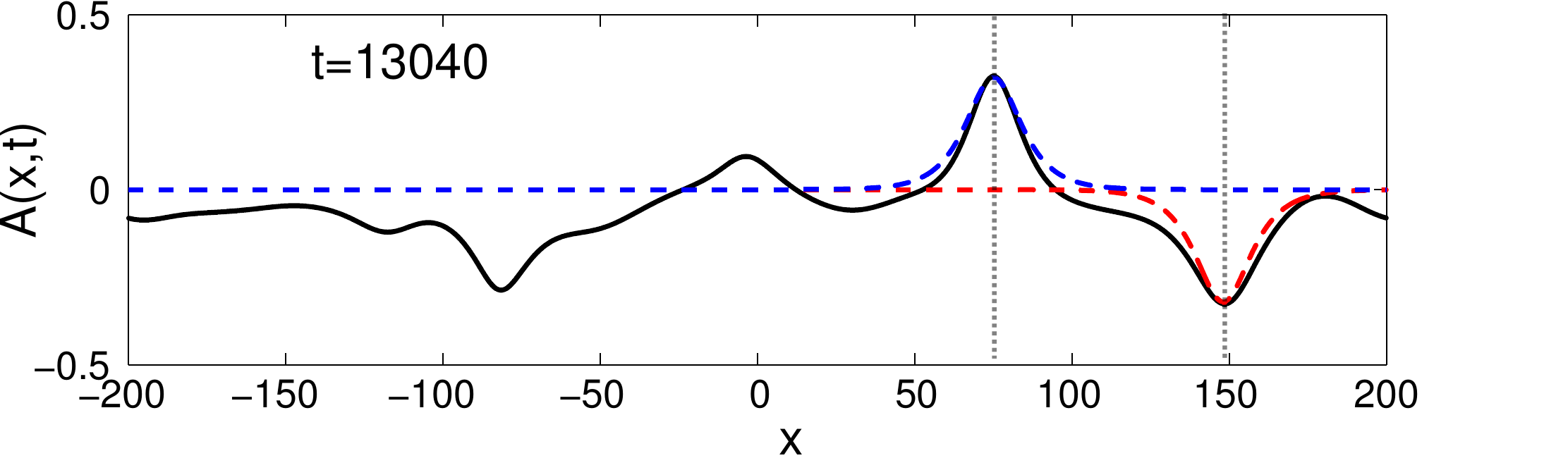}
\caption{(Color online) Top panel: Contour plot showing the energy functional $\mathcal{H}(t)$, for $q=1.5$, $Q=0.1$, $u_0=1$, $\epsilon=0.1$ for an initial condition corresponding to an unstable plane wave. Bottom panel: Profile of the gauge field $A(x)$, (solid [black] line), at $t=13040$ after the instability has set in. Dashed (blue and red) lines, correspond to the the solutions of Eq.~$(2.19)$, fitted with the parameter $u_0$ at the peak of the oscillons (indicated by the dotted [grey] lines at $x=75$ and $x=150$).}
\label{fig6}
\end{figure}

To illustrate 
the above, we have numerically integrated 
Eqs.~(\ref{eq:eq4})-(\ref{eq:eq5}), with an initial condition corresponding to the plane wave of Eq.~(\ref{plane}), 
perturbed as in Eq.~(\ref{miansatz1}), with $\varepsilon=0.1$, $W_0=1$, $Q=0.1$ (inside the instability band), 
and $\Theta_0=0$; the rest of the parameters used are $q=1.5$, $ \epsilon = 0.1$, $u_0 = 1$ and $k=0$. In the top panel of Fig.~\ref{fig6}, 
we show a contour plot of the energy density, $\mathcal{H}(t)$. It is observed that, at $t\sim 10,000$, localization
of energy is observed due to the onset of the modulation instability. This localization is due to the 
 fact that harmonics of the unstable wavenumber $Q$ of the perturbation are generated, which deform the 
plane wave and lead to the formation of localized entities. In fact, the latter are eventually reformed 
into oscillons, as is clearly observed in the 
bottom panel of Fig.~\ref{fig6}: there, we show the profile of the gauge field $A(x)$ [solid (black) line] 
at $t=13040$, and we identify at least two well formed oscillons located at $x=75$ and $x=150$.
The latter are found to be in a very good agreement with the  analytical profile depicted by the dashed (red and blue) lines. 
To perform the fitting, we plotted the approximate solutions of Eq.~(\ref{eq:eq14}), using the value at the peak of the oscillons,
 in order to identify the oscillon amplitude $u_0$. Then, 
the solution was also displaced by a constant factor, in order to match the center of the oscillon. Additionally, 
time was set to zero ($t=0$) and the fitting was made at the beginning of the period of oscillation for the particular oscillons. The 
very good agreement between the analytical and numerical field profiles 
highlights the fact that the oscillons considered in this section can be 
generated spontaneously via the 
modulational instability mechanism.      	

%
\subsection{Oscillating kinks}
\begin{figure}[t]
\center
\includegraphics[scale=0.4]{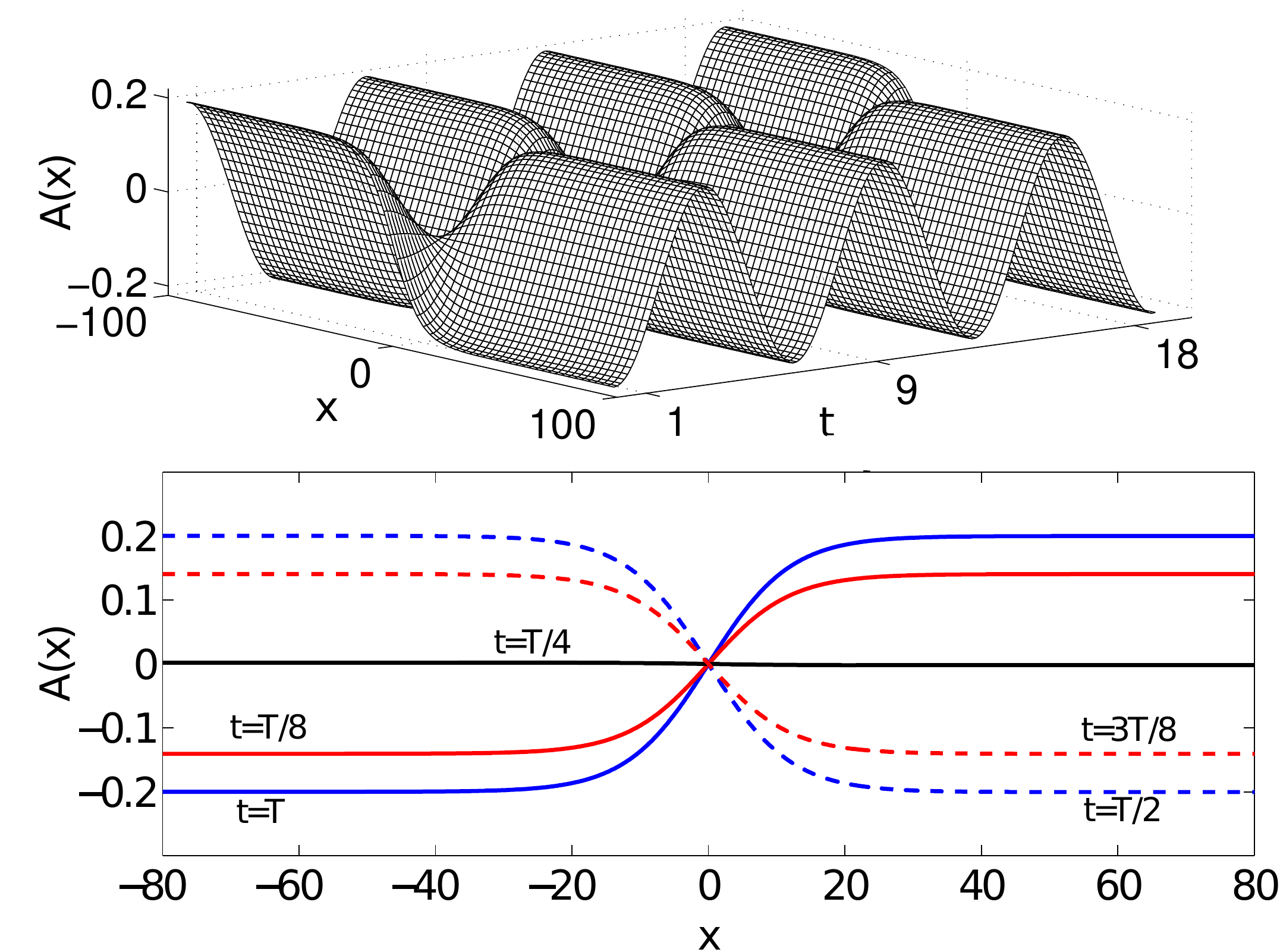}
\caption{(Color online) Same as Fig.~1, but for a kink solution, for $q=1.8$.}
\label{fig7}
\end{figure}

\begin{figure}[t]
\center
\includegraphics[scale=0.4]{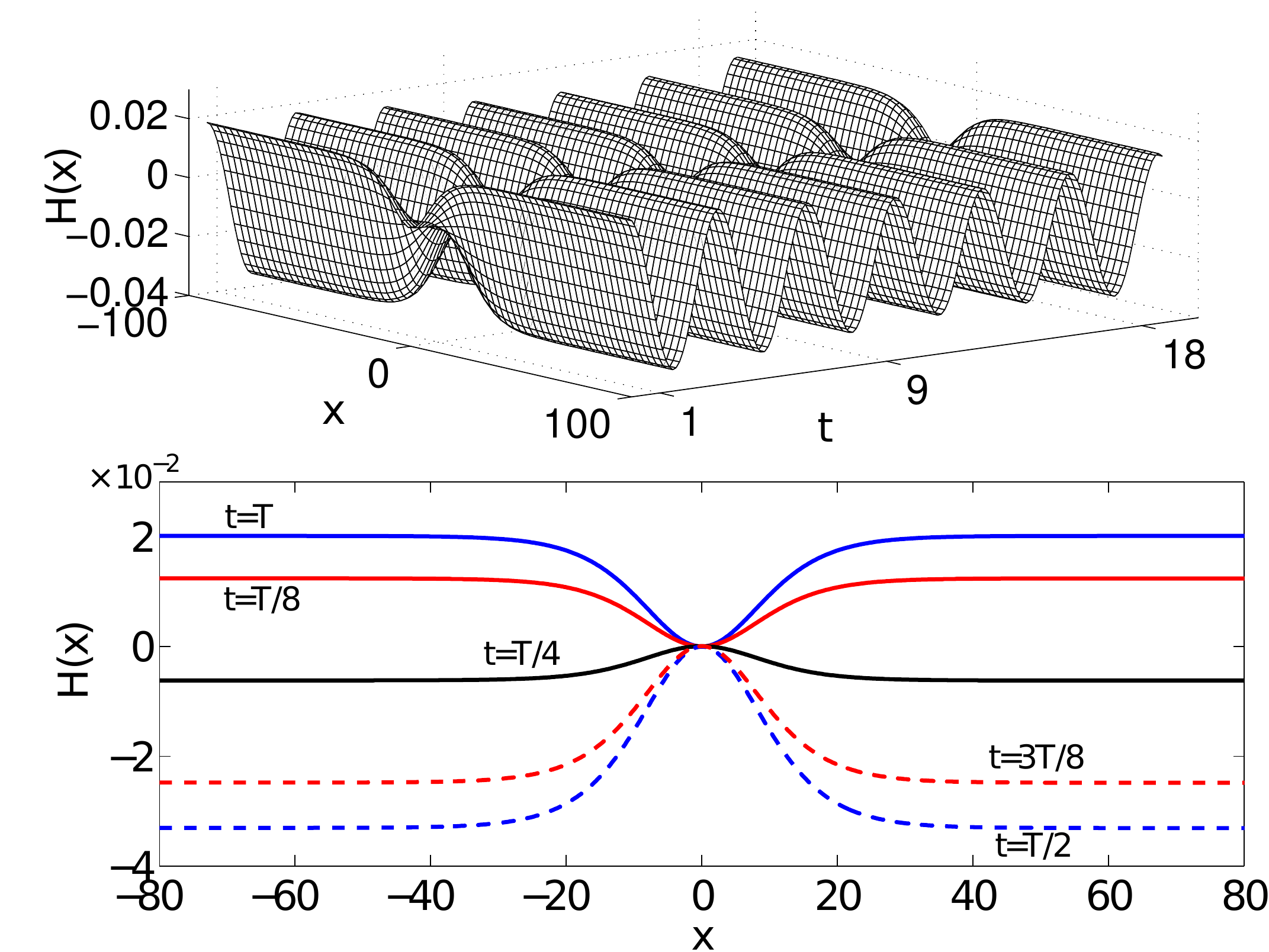}
\caption{(Color online) Same as Fig.~2 for the kink solution, and for $q=1.8$.}
\label{fig8}
\end{figure}

\begin{figure}[t]
\center
\includegraphics[scale=0.4]{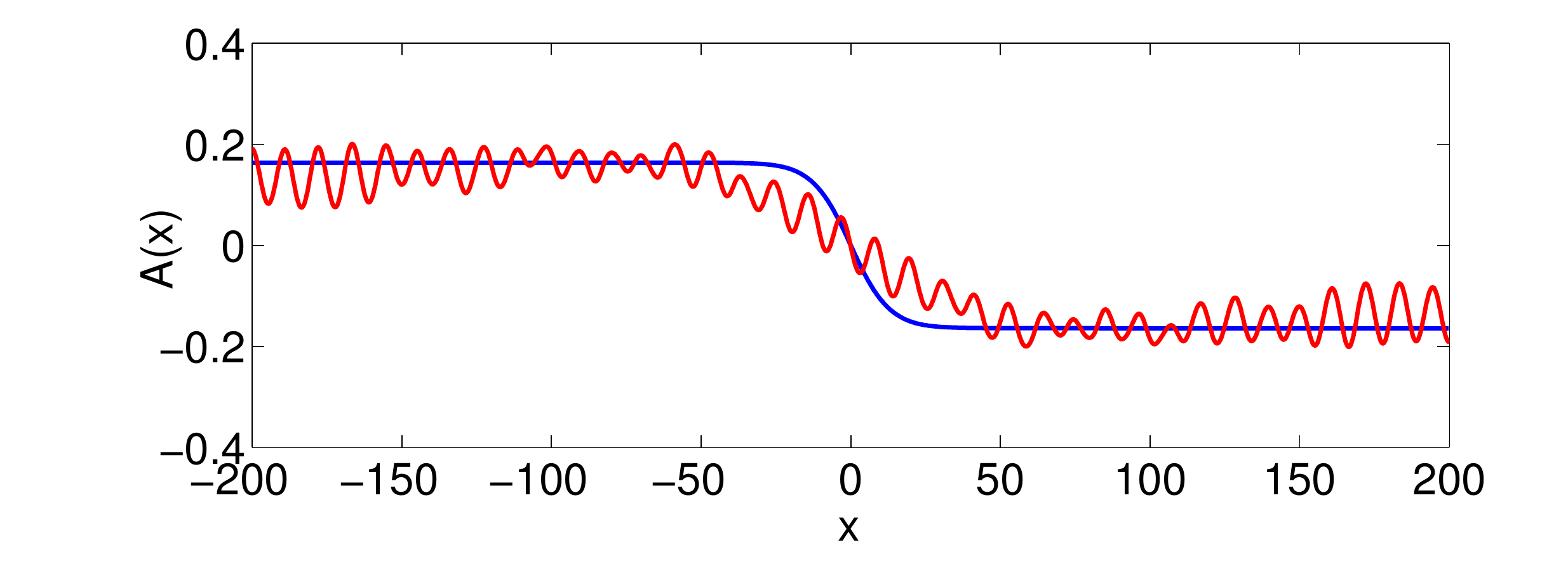}
\includegraphics[scale=0.4]{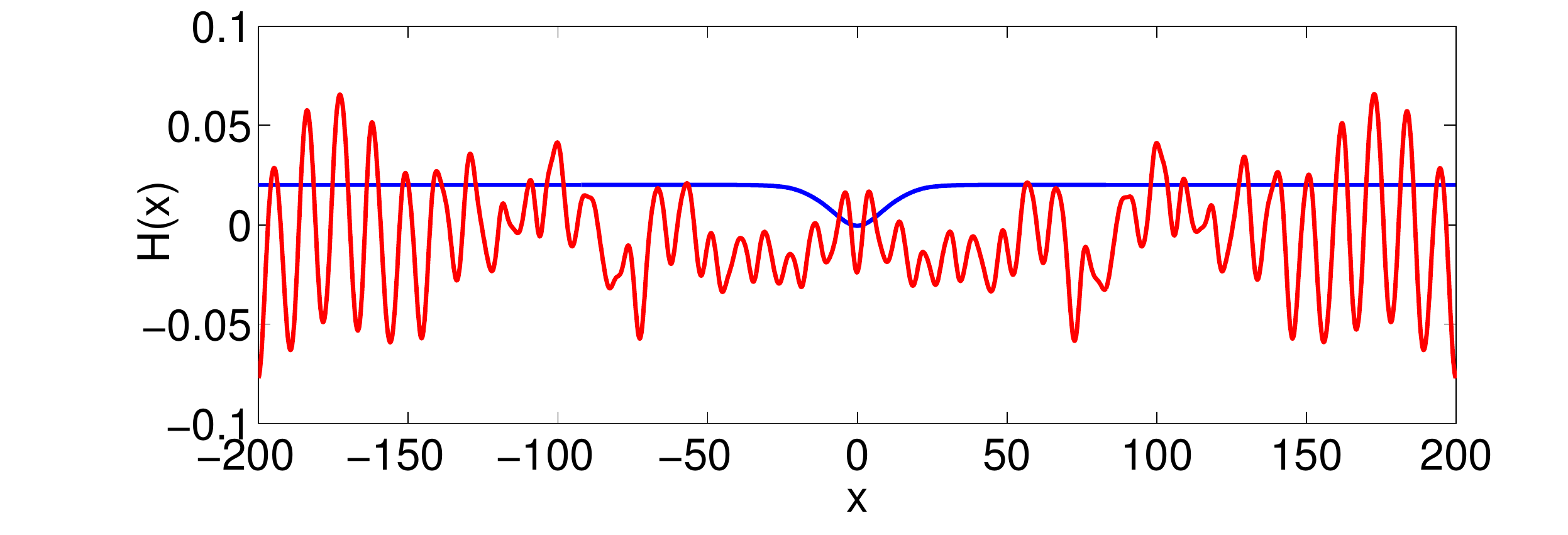}
\caption{(Color online) Top panel: profile snapshots of the gauge field at the beginning [solid (blue) line], 
and at the end of the simulation [solid (red) line], for $q=1.8$. Bottom panel: same as in top panel but for the Higgs field.}
\label{fig9}
\end{figure}

We now 
proceed with the numerical study of solutions corresponding to the kinks of the NLS Eq.~(\ref{eq:eq12}).
Such solutions in the form of Eqs.~(\ref{eq:eq17})-(\ref{eq:eq18}) are expected to 
exist in the parameter region $1.63<q<2$. Performing the same procedure as in the previous section, 
we numerically integrate the equations of motion 
and study the relevant dynamics. In the top panels of Figs.~\ref{fig7} and \ref{fig8}, the contour plot 
of a kink solution is plotted, showing the evolution of both fields $A$ and $H$, during the interval 
of three and six periods, respectively (for $q=1.8$). In addition, the profiles of both fields are shown 
in the bottom panels of Figs.~\ref{fig7}-~\ref{fig8}. The
oscillating kinks were found to be unstable for all values of the parameter $q$ within their region of existence. The
instability is manifested by an abrupt deformation of the kink solution, even near its core, characterized by the inverse 
width $l_d$, (where $l_d= 2 \sqrt{|s|}\epsilon u_0$). In Fig.~\ref{fig9}, the top panel shows the profiles of the gauge 
field at the beginning ($t=8$) [solid (blue) line], and at the end of the integration ($t=1200$) [solid (red) line]. 
While the shape of the kink is more or less preserved, it is clear that the solution profile has been significantly deformed.
More importantly, the Higgs field $H$, shown in the bottom panel, not only has been distorted but it has also 
become an order of magnitude larger than its initial amplitude.
Thus, it can be concluded that oscillating kinks are unstable.



%

\begin{figure}[t]
\center
\includegraphics[scale=0.4]{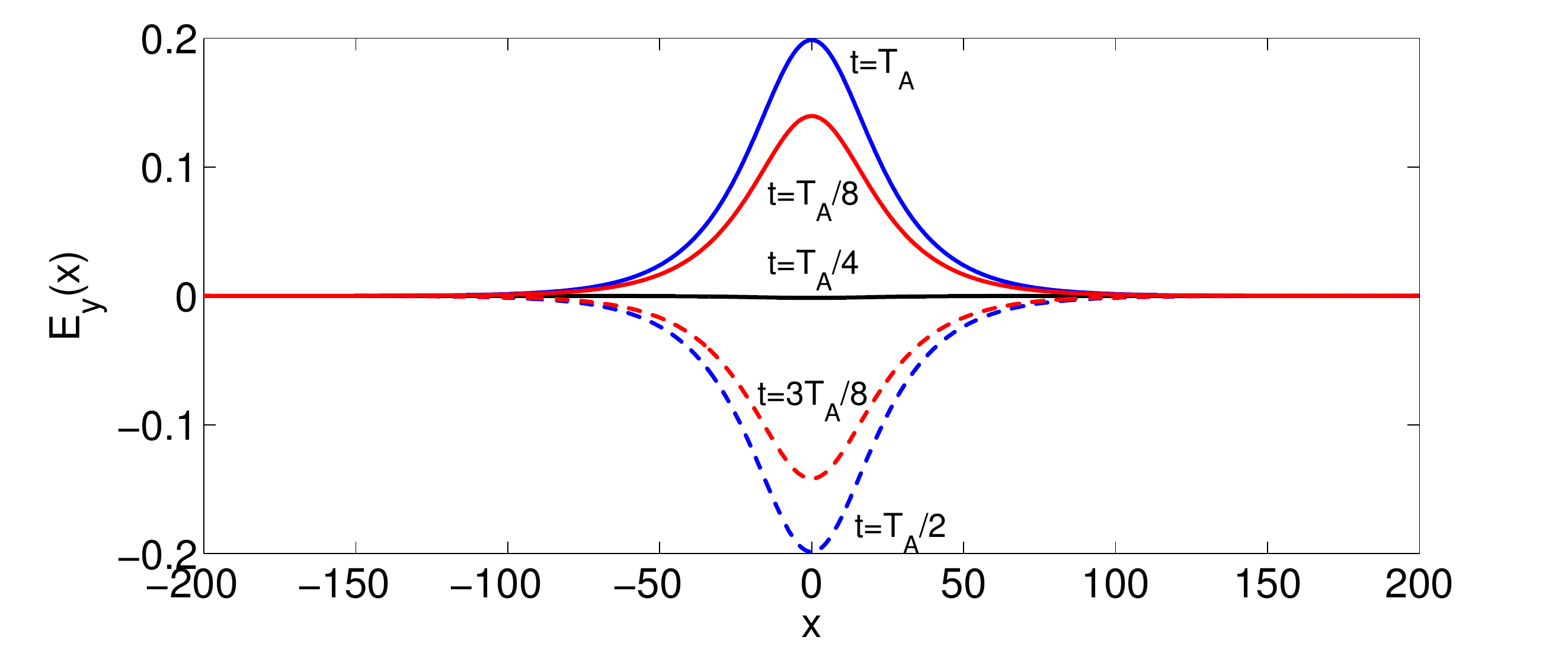}
\includegraphics[scale=0.4]{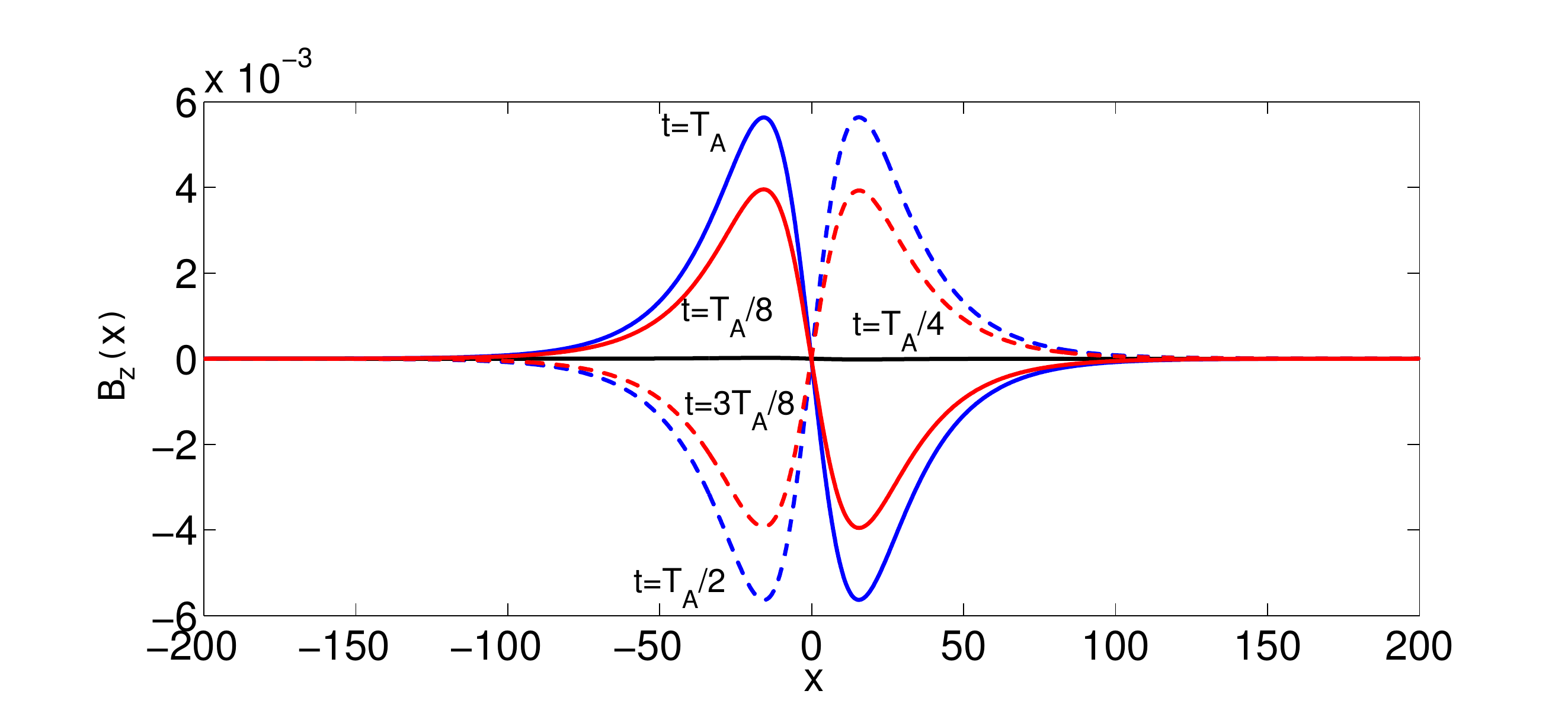}
\caption{(Color online) Top panel: profile snapshots of the electric field $\mathcal{E}_y (x)$ for $q=1.5$. Bottom panel: profile snapshots of the magnetic field $B_z (x)$ for the same $q$.}
\label{fig10}
\end{figure}
%


\section{Discussion and conclusions}

In this work we have presented a class of classical solutions of the 1D Abelian Higgs model obtained with the use 
of a multiscale expansion method. 
The key assumption in our treatment is that the scalar field amplitude is much smaller than that of the gauge field; 
the relevant ratio was then used as a formal small parameter in the perturbation expansion. 
We have shown that the equations of motion can be reduced to a NLS equation; by means of the latter, 
localized solutions in the form of oscillons and oscillating kinks were derived.
Results  by  numerical integration 
of the original 
equations of motion where found to be in very good agreement with the analytical findings.
In addition, we have numerically studied stability of the solutions against 
a Gaussian noise with amplitude 
up to $10\%$ of the Higgs field amplitude, and found that the oscillons remain robust.  The robustness of oscillons is also complemented by more recent findings \cite{pre} where it is indicated that only oscillon solutions for the gauge and the Higgs field are long-lived. This leads us to the conclusion that oscillons dominate in the solution space of the Abelian-Higgs model.

It is also relevant 
to discuss the possible connection of the 
presented solutions to the physics of superconductors. 
One could, in principle, write down the form of the magnetic and electric fields originating from the gauge field 
$A$ which was chosen to be in the $\hat{y}$ direction:
$\vec{B}(x,t)=\partial_x A(x,t)\hat{z}$, and $ 
\vec{\mathcal{E}}(x,t)=-\partial_t A(x,t)\hat{y}$.
 
In the case of the oscillon solutions given in Eqs.~(\ref{eq:eq14})-(\ref{eq:eq15}), the above equations describe the electric field in the $y$ direction, which produces a magnetic field in the $z$ direction; 
both fields are localized around the origin of the $x$ axis -- cf. 
Fig.~\ref{fig10}, where the profiles of the fields are shown.
This can be thought of as a configuration, describing 
Superconductor--Normal metal--Superconductor (SNS)
Josephson junction~\cite{paterno}, where two superconductors are linked by a
thin normal conductor placed at the origin. Then, our solutions describe a condensate $\phi=(\upsilon+H(x,t))/\sqrt{2}$ that oscillates around its vev near the origin, and acquires its vev value when entering the superconductors. Accordingly, 
the magnetic field is shown to oscillate inside the normal conductor, but vanishes exponentially inside the superconductors 
as per the Meissner effect.



Our approach not only reveals a new class of solutions of the Abelian-Higgs model, but 
also dictates a straightforward
general strategy for the search of non-trivial dynamics in models involving classical fields with 
nonlinear interactions. In particular the dynamics of such models is governed by the nonlinear Schr\"{o}dinger equation. It 
is an interesting perspective to determine the impact of these solutions on the thermal and quantum behaviour of the
involved fields. 
Such studies is an interesting theme for future work.

{\bf Acknowledgments.} Illuminating discussions with L. P. Gork'ov are kindly acknowledged.

\end{document}